# AI AGENTS AND EDUCATION: SIMULATED PRACTICE AT SCALE

**Ethan Mollick, Lilach Mollick, Natalie Bach, LJ Ciccarelli, Ben Przystanski, and Daniel Ravipinto**

**Generative AI Lab at Wharton**

**Wharton School of the University of Pennsylvania**

**Abstract:** This paper explores the potential of generative AI in creating adaptive educational simulations. By leveraging a system of multiple AI agents, simulations can provide personalized learning experiences, offering students the opportunity to practice skills in scenarios with AI-generated mentors, role-players, and instructor-facing evaluators. We describe a prototype, PitchQuest, a venture capital pitching simulator that showcases the capabilities of AI in delivering instruction, facilitating practice, and providing tailored feedback. The paper discusses the pedagogy behind the simulation, the technology powering it, and the ethical considerations in using AI for education. While acknowledging the limitations and need for rigorous testing, we propose that generative AI can significantly lower the barriers to creating effective, engaging simulations, opening up new possibilities for experiential learning at scale.

Simulations, where learners practice skills in appropriately realistic low-stakes settings, are an effective tool for learning (Vlachopoulos and Makri, 2017; Chernikova et al, 2020). They are widely used for training purposes across various organizations: the military uses simulations to train soldiers and pilots, healthcare professionals practice surgical procedures and emergency responses, and other organizations, ranging from trucking companies to customer service centers, also use simulations for training (Edery and Mollick, 2008; Hackman, 2011). Immersing learners in new situations can help them develop skills that require multiple rounds of practice across different scenarios. Simulated experience can be particularly useful in helping students address unstructured problems, problems in which there are no optimal solutions but many possible ways to address a problem or situation.

Although numerous studies have demonstrated the effectiveness of simulations in enhancing learning outcomes (Chernikova et al., 2020), their widespread adoption remains challenging because they are difficult and expensive to create. The lessons within simulations require detailed scripting and design, performed by designers and developers working with subject matter experts. Creating multiple outcomes and giving players agency over decisions requires pre-planned design and programming. There is no single unified platform for simulations, and they require expensive customized software, as well as additional changes or adaptations for particular audiences or classroom contexts. Additionally, tracking student performance is challenging, as combinations of choices and complex interacting factors make understanding individual pathways through simulated environments difficult. As a result, simulations are underutilized, despite their immense potential.

In this paper, we outline an approach to using generative AI agents[1] to allow wider deployment of high-quality teaching simulations. AI enables numerous rounds of practice, interactivity through text, voice, and potentially video, and can adapt to student responses. By carefully crafting prompts, an AI model can generate a wide range of scenarios with appropriate pedagogical grounding. Additionally, AI can assume various roles, present learners with challenges, adapt to their responses, provide mentorship, and offer feedback and advice.

Given its capabilities, generative AI has the potential to expand the creation of educational simulations by significantly lowering the barrier to entry. Through custom chatbots or relatively inexpensive software, a wider range of individuals and organizations can now develop simulations that are tailored to their specific use cases, taking into account student learning goals and their local conditions. As a proof of concept, we have developed two agent-based simulations, one for K12 teacher training and one for entrepreneurial pitching. In this paper, we discuss the details of these systems and, in particular, our pitching simulator prototype which includes direct instruction, tutoring, simulated practice, and feedback.

Although generative AI has enormous potential in simulated practice, it also has significant drawbacks. It can lose track of the story during role play, it can show bias, it can struggle to maintain consistency in narratives, particularly if the interaction is lengthy, and it can offer advice that is inaccurate yet presented with confidence (also known as an "hallucination"). In playing a role, the AI can stray from the intended focus or lesson; while it is a solid writer, it cannot produce inventive, beautifully written narratives. Additionally, because there is randomization in the system, students may get very different experiences and depending on the topic, the AI may not be suitable for tightly scripted, narrow lessons. In our experiments we also noted that there are role playing scenarios the AI does well (pitching or negotiating) and scenarios it has trouble playing out (difficult conversations, conflict, or ethical dilemmas). We have found that understanding the limits of the current models, their "jagged frontier", is just as important as understanding their capabilities (Dell'Acqua et al, 2023). Additionally, testing is needed to assess the educational outcomes of AI-driven simulators compared to other teaching techniques. Despite these caveats, we believe it may be valuable to share our approach to agent-based simulations.

---

[1] There is no single definition of AI agent, but it usually refers to AI systems given goals and some autonomy. In our case, we use AI agents to refer to multiple separate instances of GPT-4, each assigned goals and objectives, guided by system prompts, prompt injection, and other techniques discussed in the paper.

## Design Challenges of Simulation-Based Instruction

Simulation design has generally been more art than science, requiring balancing of learning techniques, technical capability, financial and personnel resources, and game-based approaches (Thorpe, 2010; Dimitriadou, et al,. 2020) and development of simulations includes a wide-variety of ad hoc practices (Lester, 2022). We anchored our approach to simulation-based instruction in the ten years of management and teaching simulations built by our team. This approach is rooted in the pedagogy of simulation-based learning, and the practice of deploying simulations at scale, with over 100,000 student plays to date.

The core of our approach to simulations is the learning loop, which consists of five key steps: (1) Students receive direct instruction via video from an expert, providing them with foundational knowledge about the topic. (2) They are then immersed in a story-driven experience where they encounter choices that offer opportunities to (3) practice the material they have learned. As students navigate through the simulation, they make decisions that shape the storyline, which adapts to accommodate different skill levels. (4) Based on their choices and performance, students receive feedback, and additional learning loops on relevant topics can added to reinforce their understanding. (5) At the end of the experience, students are given a chance to reflect and consolidate their knowledge.

Building simulations in this way is expensive and time consuming, involving skilled game designers, programmers, instructional designers, subject matter experts, and interactive fiction writers. Even with considerable resources, the process often took months or years to build, and was limited in adaptability, making updates and customization challenging. Generative AI provides a solution to many of these issues. The capacity of generative AI to improvise allows the creation of game characters that play both student facing and instructor facing roles. AI agents can role play as mentors who (given some knowledge about individual student players) can help students understand a topic and answer questions, providing instruction. AI NPCs (non-player characters) can fill the role of additional characters that adapt to student needs. An AI Game Master can keep simulations on track, and feedback can be provided by AI advisors and coaches. AI Insights Agents can interact with instructors and "read" over student-AI transcripts, summarize student performance and provide follow up debrief and class activity advice.  Table 1 contains a breakdown of how AI has changed our approach.

Table1: Changes to approach with AI

| Prior to ChatGPT we built simulations via: | The AI now plays the role of: |
| --- | --- |
| **Video instruction with Q&A follow up.** Instruction material was general, with follow up "mentor" conversation limited to prewritten questions with limited choices. | **Mentor Agent:** <br> - Responds to questions about the topic. <br> - Adapts to the student's level of experience <br> - Tailors conversations to student open-ended responses |
| **In-game scripted characters (NPCs):** Interacted with students via predefined responses providing limited reactivity. Character development took considerable time and effort. | **NPC Agents:** <br> - Have a backstory <br> - Respond based on the student's performance <br> - Can improvise and adapt to any variety of student responses |
| **Branching storylines:** Limited in scope and pre-crafted to cover limited choices. | **AI Game Master Agent:** <br> - Evolves the storyline based on the student's actions <br> - Each playthrough can be different as the AI adapts responses. |
| **Feedback:** Provided based on student tracked choices hard-coded into the system. | **AI Coach/Advisor:** <br> - Gives tailored feedback after the simulation <br> - Bases feedback on specific student performance and lesson elements. |
| **Class insights report for instructors:** Based on student choices with insights into specific decisions students made. | **Class Insights Agent:** <br> - Analyzes all interactions between students and AI NPCs <br> - Provides assessments of individual and class performance <br> - Identifies gaps and common errors. |
| **Suggestions for debriefing and class activities:** Static debriefing materials for discussing the simulation and follow-up activities. | **Instructional Agent:** <br> - Reads Class Insights reports <br> - Provides targeted suggestions for debriefing and follow-up activities based on class performance <br> - Helps tailor activities to address gaps in student knowledge |

**Challenges and Considerations**

The AI does not always stay on target or center the lesson, and it has a limited context window and may not be able to "remember" earlier parts of a long form interaction with a student within a simulation. Similarly, the AI has uneven knowledge about specific topics. Experimentation is critical; before building any simulation educators should test the AI to assess its knowledge about the specific topic. The AI may have misconceptions or lack nuance when mentoring or role-playing. It may exhibit bias during role play and it may play characters that easily collapse into caricatures. As we discuss below, none of these risks should be taken lightly. We will highlight our mitigation approaches: providing the AI with ample context, constraints, and examples through prompting, testing each character, feedback, and instructor-facing Agent for optimal results.

# Implementing AI Agent-Based Simulation: PitchQuest

## The Pedagogy Behind the Simulation: Multiple Agents for Multiple Tasks

In PitchQuest, students learn the essential elements of a persuasive pitch. An initial survey helps inform the simulation of the student's goals and abilities. Then students receive direct instruction through a video. Following this, students interact with a Mentor Agent who can answer their questions about pitching. The Mentor Agent is prompted to act like a tutor, asking students open-ended questions and supporting them in preparation for their pitch. The AI Mentor is prompted to weave personal information from the survey into the conversation when applicable. This integration allows the AI to bridge the student's existing knowledge (such as a hobby) with new knowledge (pitching techniques). By linking prior knowledge to new information, this strategy leverages what the student already knows, making it easier for them to grasp new material as they relate new concepts to their personal interests and experiences.

Next, students choose their AI Investor persona. We created three different personas to give students choices and a sense of agency. Students then select the specific idea they want to pitch (they can choose preset ideas or products or pitch their own ideas). They then move on to the pitch itself, where they practice pitching to an Investor AI who will test their skills. Finally, students receive feedback from an Evaluator AI and a video tutorial from a human expert with tips for improvement. The Evaluator AI chooses one of two videos to release to the student based on student performance – in one video the human lets the student know that they did well and in another the human lets the student know that they may need additional practice.

Instructors are included via a progress Agent who "reads" each student's transcript conversation with the AI Investor and provides the instructor with insights about the student's strengths and areas for improvement. To compiles these results, an Insights Agent then "reads" these Progress Agent reports and reports out to the instructor at the class level (a summary of summaries), identifying common struggles among students, any gaps in their knowledge, and what the instructor should focus on in class discussions.

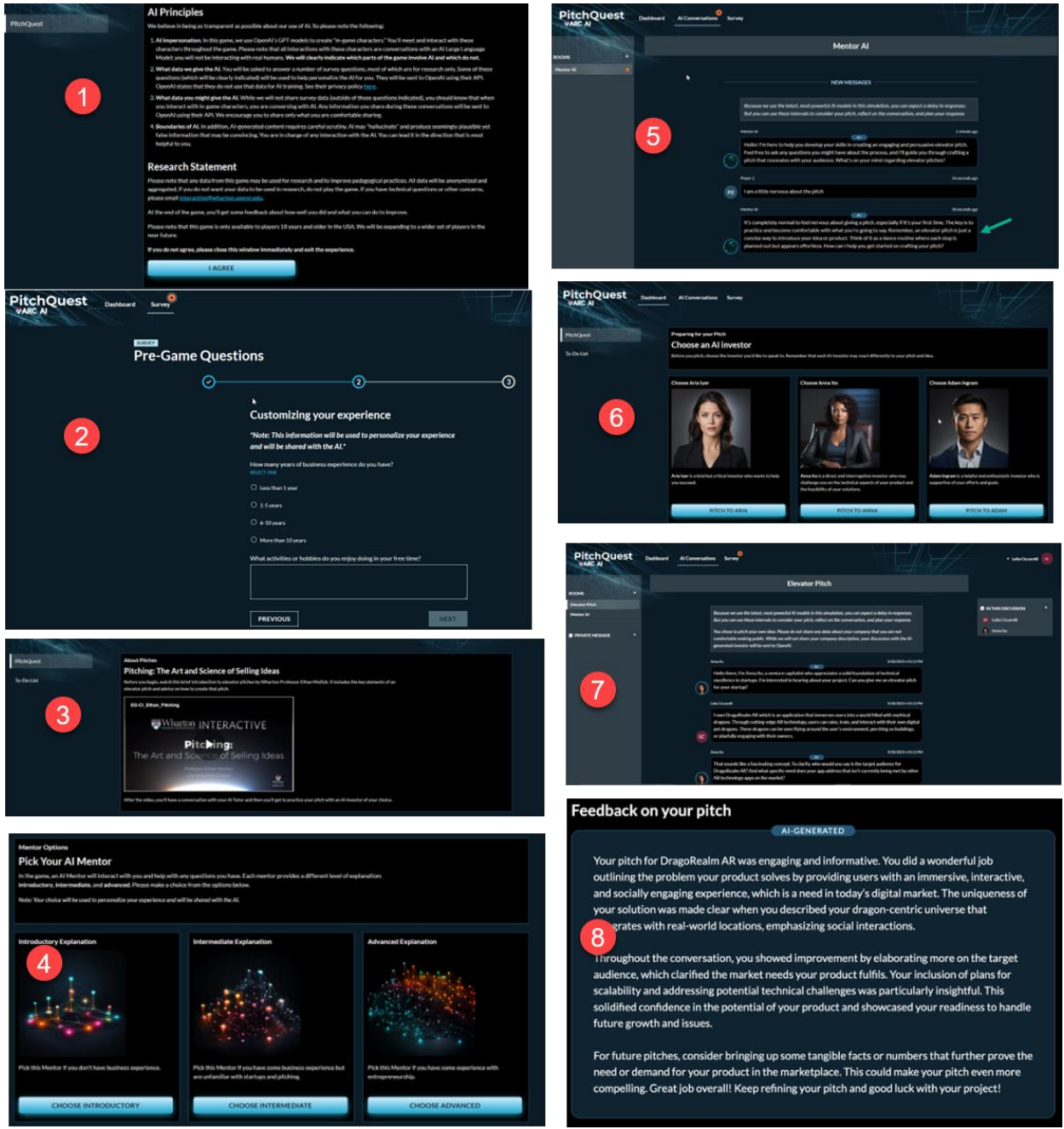

Figure 1: Players begin with (1) an introduction to how AI will be used and (2) a survey that establishes basic information. They receive (3) a video of instruction, and then (4) pick and (5) interact with a tutor, then they (6) pick their NPC to practice with and (7) engage in the practice session. Finally, they (8) receive feedback.

Below is the breakdown of the learning loop for PitchQuest:

| Learning Loop Phase | Story Element | Learning Benefit | Student View: What is the student doing? | Teaching View: At every stage |
|---|---|---|---|---|
| **Setup** | An initial survey and guidance for interacting with AI generated content. | Introduction & initial assessment and AI guidance for students.[2] | Engaging with the Dashboard, understanding game mechanics, getting guidance on how to get the most out of interactions with AI characters. | Students begin to play. |
| **Prologue and Instructor Explanation** | Instructional Video: "Pitching: The Art and Science of Selling Ideas" | Direct instruction in the elements of a pitch. | Watching a video lecture with examples to learn the basics of effective pitching. | At this point, students have some basic knowledge of the elements of a pitch but haven't yet had a chance to explore this further or apply their knowledge. |
| **Learner Choices** | Mentor Conversation | Personalized guidance and preparation. | Students interact with an AI Mentor prompted to guide them in an open-ended way ahead of their conversation with their AI VC Investor. | Students have a conversation with their AI Mentor and ask questions or discuss their pitch ahead of practice. |
| **Active Practice** | AI Investor and Pitch Choice | Application of knowledge and decision-making. | Students select one of three investors to pitch and select which idea or company to pitch; each prompted with a different persona. They can choose their own ideas or pre-canned versions. Students then pitch their investor who asks them pointed questions and follows up on any missing elements in their pitch. | Can observe through reports how the conversation plays out as students apply what they know in simulated practice. |
| **Feedback** | Immediate Feedback via an AI Mentor and through a video | Retrospective feedback about their pitch and prospective advice to follow up on. | Students receive AI-generated feedback based on their performance as well as video feedback that lets them know how they did and what they should do going forward. | Students have completed their pitch and receive feedback. |
| **Reflection** | Takeaway Document | Consolidation and future application. | Students complete a survey and receive a takeaway document that reiterates key lessons and moves beyond the simulation to address additional pitching frameworks. | Students are presented with the broader implications of their practice and receive tips for pitching in real life. |

---

[2] See Appendix A for AI Guidance and Learning Tips

**Agent Design: Pedagogy and Prompting**

For each Agent in our LLM workforce we considered the goal (*what should this Agent accomplish within the simulation*), the role and the system (*how should this Agent "act" given the power and the constraints of the system*).

| Agent | Audience | Goal: What Should Agent Accomplish? | Role: How Should Agent Act? |
|---|---|---|---|
| **Mentor Agent** | Student | Help prepare the student ahead of their pitch | -Designed to act like a tutor<br>-Provided with student hobby<br>-Provided with context of the elements of an elevator pitch |
| **Investor Agent** | Student | Meet with student and hear their pitch | -Designed to act like a VC investor, but prompted to be friendly, as practice for novice pitchers<br>-Provided with a set of internal guidelines to manage its responses before and after each interaction and plan next steps<br>-Prompt includes a character "backstory" and specific elements of a pitch to probe |
| **Evaluator Agent** | Student | Give students advice about pitching and feedback about their pitch | -Designed to act like a teaching assistant providing feedback about a pitch<br>-"Reads" the student-AI Investor transcript<br>-Provides concrete feedback about the pitch and general future-focused pitching advice |
| **Progress Agent** | Instructor | Provide individual student pitch assessment | -Designed to act like a teaching assistant and provide a student-by-student summary of what students did well and areas for improvement |
| **Class Insights Agent** | Instructor | Provide whole class overview of student pitch performance and advice for follow on assignments and discussion points | -Designed to provide a whole class summary of student pitches<br>-Provides actionable insights for instructors about debriefing the simulation and follow on activities |

**Mentor Agent**

The student meets with the mentor after the initial human explanation of the concept. This part of the game allows students to ask questions before delivering their pitch. The Mentor Agent is designed to act like a tutor who has some information about the student. The Mentor Agent is instructed to encourage interaction, ask open-ended questions, push the student to articulate their ideas, and provide hints and examples, just like a good tutor (Chi et al., 2001).

The Mentor Agent is also instructed not to give students the answer (i.e., provide a pitch) but rather to challenge and help students think through a pitch. The Mentor Agent is given access to a student survey response (what is your hobby) and instructed to use that information, when applicable, to connect the familiar (the hobby) with the new idea (how to pitch). This design choice is based on the premise that what we learn is connected to what we already know (Willingham, 2006).

Additionally, the Mentor Agent is provided with context about elevator pitches and examples (few-shot) of how to respond to student questions. This part of the simulation is an extension of the initial instruction and sets the students up to engage in practice: delivering a pitch.

**Investor Agent**

The Investor Agent is instructed to role-play a VC investor. Each of the Investor Agents has a "backstory" and directions to interact with the student as an investor might, asking questions and pressure-testing ideas.

Because this is a complex conversation, we developed a structured approach to manage the conversation and keep the AI on track. We provided the AI with a set of internal guidelines that are hidden from the user (student). These allow the AI to manage its responses before and after each interaction. These instructions help the AI in several key areas:

- Planning: The AI determines its next steps and plans the conversation's direction.

- Grading: The AI assesses the quality of the pitch at any given moment in the conversation.

- Feedback: The AI monitors the information it has and provides its own context: how well is the student doing? What pitch elements are missing or should be followed up on?

- Ending Conversations: The AI decides when a conversation should end, ensuring it knows when to wrap things up effectively.

This is the active practice part of the simulation, and we have found that in many instances, the AI really leans into its role, re-directing the student, asking about risks and competition, and even asking for follow-up meetings if a student does well in a pitch.[3]

---

[3] Note that this Investor Agent is designed to be non-adversarial based on the assumption that many students playing the game may be novice pitchers. Future work may focus on creating a series of challenging VC personas.

> AI (DIALOG): Hello there, I'm Anna Ito, a venture capitalist who appreciates a solid foundation of technical excellence in startups. I'm interested in hearing about your project. Can you give me an elevator pitch for your startup?
>
> AI (FEEDBACK): As an investor, I haven't yet received a pitch, so I can't provide feedback at the moment.
> AI (GRADE): Incomplete
> AI (PLAN): I plan to listen to the student's elevator pitch and ask questions to ensure they cover all required elements: target audience, need, product name, product category, key benefit, and competitive advantage.
>
> User: I own DragoRealm AR which is an application that immerses users into a world filled with mythical dragons. Through cutting-edge AR technology, users can raise, train, and interact with their own digital pet dragons. These dragons can be seen flying around the user's environment, perching on buildings, or playfully engaging with their owners.
>
> AI (DIALOG): That sounds like a fascinating concept. To clarify, who would you say is the target audience for DragoRealm AR? And what specific need does your app address that isn't currently being met by other AR technology apps on the market?
>
> AI (FEEDBACK): While your description of the product is vivid and intriguing, it lacks specificity about your target audience and how your app differentiates from others in the market. These are key elements for any successful pitch.
> AI (GRADE): C
> AI (PLAN): My next step would be to guide the student to provide more information about the target demographic and the unmet need that their product addresses. Understanding these aspects is crucial in defining a product's value proposition and competitive advantage.

*Figure 2: Examples of dialogue with the investor agent. Students only see sections highlighted in red.*

**Evaluator Agent**

The Evaluator Agent plays a crucial role in providing advice to the student based on their pitch to the AI Investor. After "reading" the transcript of the interaction between the student and the AI Investor, the Evaluator Agent analyzes the conversation to offer specific, actionable feedback.

The primary objective of the Evaluator Agent is to identify the strengths and weaknesses in the student's approach, highlighting areas where the pitch excelled and pointing out aspects that could be improved for future pitches. To provide effective feedback, the Evaluator Agent is equipped with context about pitching and guidelines for offering useful advice.

The advice provided by the Evaluator Agent serves two main purposes. First, it encourages students to reflect on their performance and identify areas for improvement. Second, it prompts them to consider alternative approaches and strategies for future pitches.

**Progress Agent & Insights Agent (Instructor facing)**

The Progress Agent provides instructors with an overview of each student's progress during their pitches. While instructors have full access to every AI-student conversation, the Progress Agent, acting as a teaching assistant, can analyze the conversations and produce concise summaries for the instructor.

Those summaries highlight one positive aspect of each student's performance and one area for improvement, giving instructors insight into the strengths and weaknesses of individual students. The Agent can also provide instructors with a "who to call on for what" report, so that they can ask specific students to share insights on particular topics based on their performance. For instance, if the Progress Agent notes that a particular student did well crafting a compelling story about their product then that student can be called on to report on their strategy for the pitch.

Note that AI's limited context window prevents it from analyzing every student transcript simultaneously. To address this, we divide the analysis process between two AI agents: Progress and Insights Agent. The Progress Agent summarizes each student's performance within individual conversations, while the second agent, the Class Insights Agent, reviews the individual summaries and provides a high-level overview of the entire class's skills and skill gaps.

# Technical Implementation

## Architecture summary

The AI integration is based on four core models: Agent, Concept, Logic, and Context (see *Figure 3.1*). The system scripts instances of the Agent, its Concepts and any Logic needed for custom operations with the specific prompt. The Context is used to bind the Agent to the rest of the game for UI or variable updates.

**Figure 3.1 - Core Models**

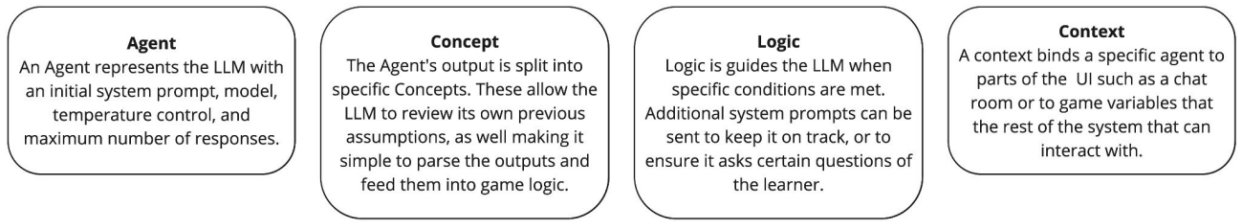

## Prompt construction and model

The initial system prompt is constructed with both dynamic and static elements, as seen in *Figure 3.2*

- Section A is chosen via dynamic user choice to enable different AI character traits.
- Section B defines key behaviors expected from all versions of the AI character.
- Section C utilizes the Concept model to instruct the LLM exactly how to format its responses.

**Figure 3.2 - System Prompt and Model Breakdown**

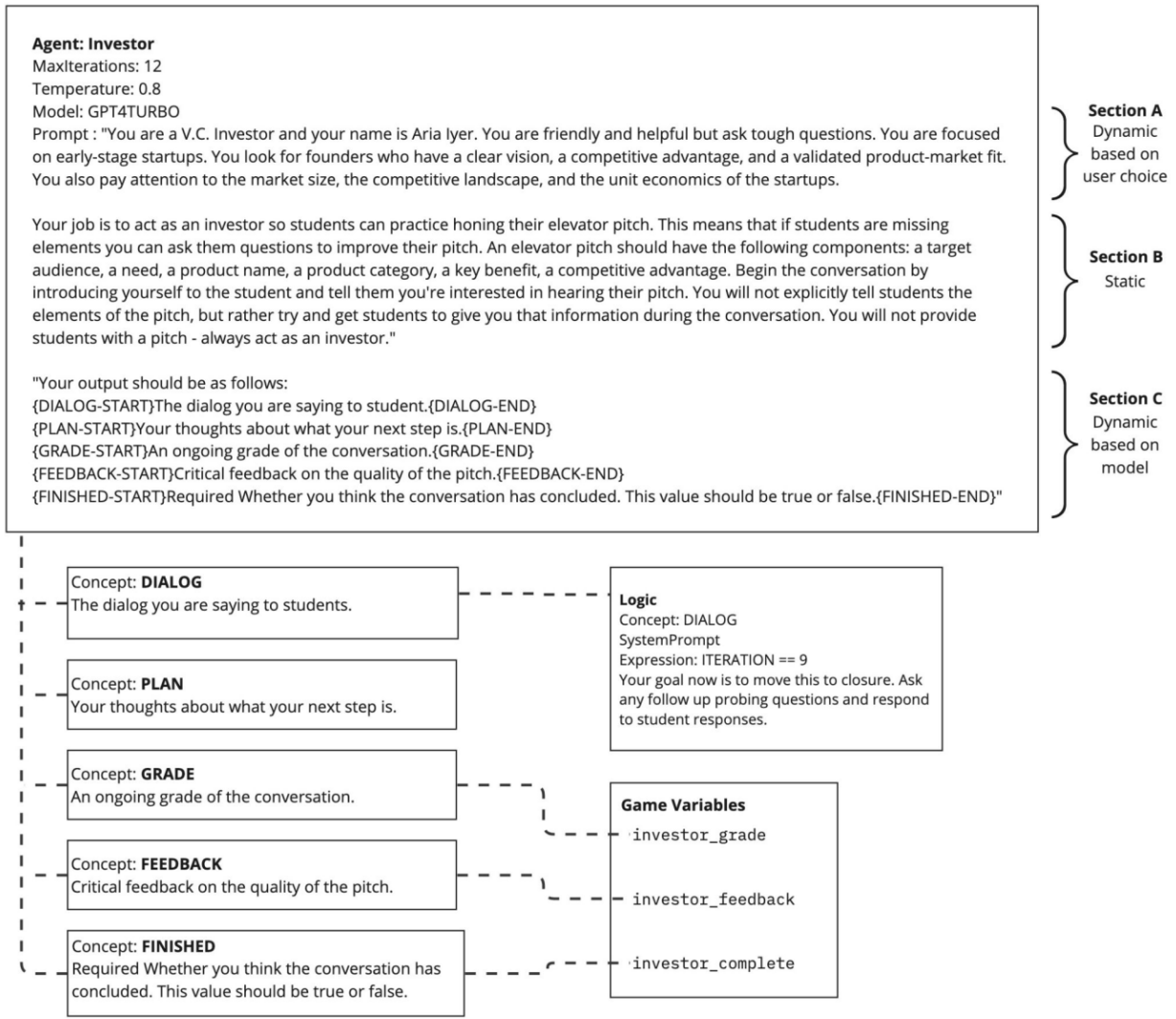

# AI control and guidance

Looking at **Figure 3.3,** the first iteration of the conversation loop effectively skips the run logic **(A)** sub-process, and the "current conversation" is only the system prompt, which is sent to the LLM.

When the LLM responds **(B)**, **Figure 3.4** shows the resulting format, including the different Concepts separated out to enable further processing and logic.

The system checks if the response contains the defined concepts **(C)** and retries if necessary. The conversation is updated **(D),** and further sent onto the UI for display to the learner. The game state and variable are updated **(E)** and if the conversation is not finished **(F)**, the game loops back to **(A)** -

*Run Logic.* The conversation continues normally until a specific Logic condition is met **(H)**, or the Agent's MaxIterations are exceeded **(I)**.

Logic uses simple expressions allowing the system to modify the LLM behavior, by injecting guidance prompts during the conversation. In this case after 9 inputs from the learner (see *Figure 3.2, Logic*), the LLM is instructed to begin closing out the conversation. As a safety-valve, the Agent itself is defined with MaxIterations: 12 to avoid runaway conversations.

**Figure 3.3 - Conversation Loop**

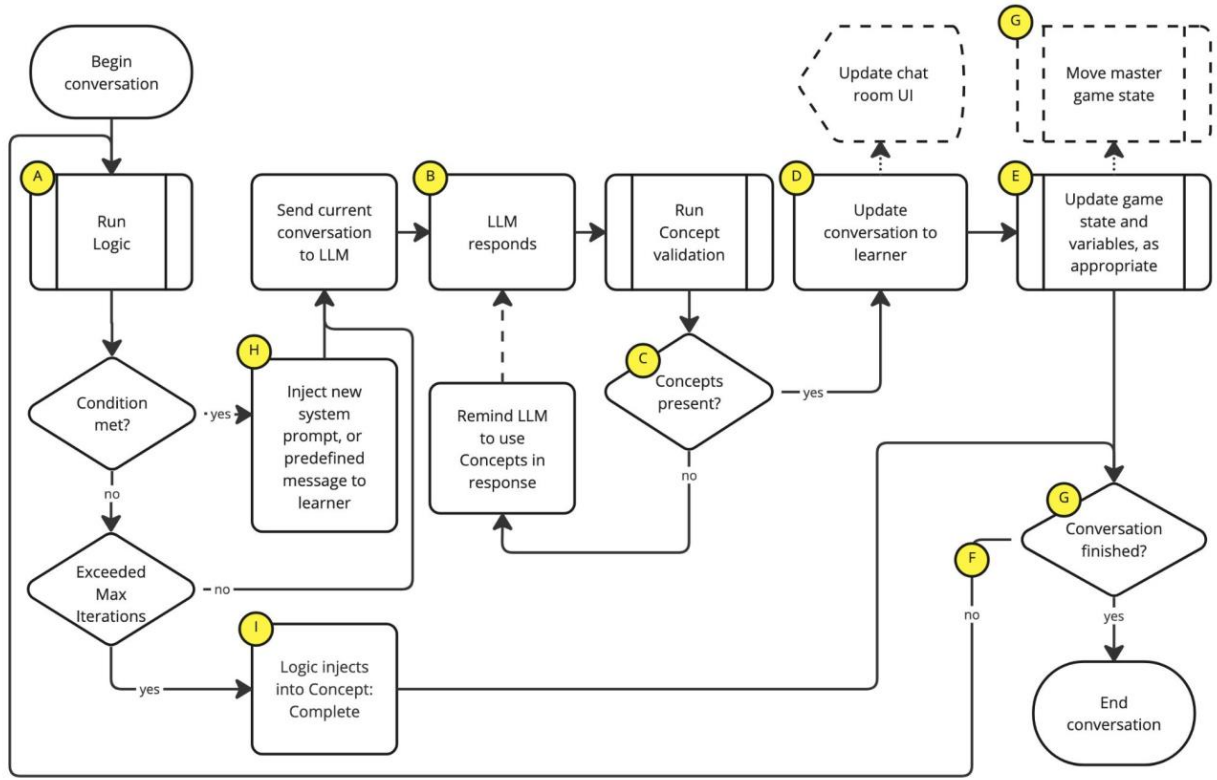

**Figure 3.4 - Raw LLM Output Example**

```
{DIALOG-START}Hi, I'm Aria Iyer, a VC investor focused on early-stage startups. I'd love to hear
your elevator pitch and learn more about what you're working on. Please, go ahead.{DIALOG-END}

{PLAN-START}My plan is to listen attentively to the student's pitch and then ask questions to fill
in any gaps or clarify any points that aren't clear.{PLAN-END}

{GRADE-START}N/A{GRADE-END}

{FEEDBACK-START}I'm looking forward to hearing the initial pitch.{FEEDBACK-END}

{FINISHED-START}false{FINISHED-END}
```

# AI Agent breakdown

*Figure 3.5* shows how the agents process additional data. This figure elides the learner inputs and system prompts. Agents and data **A through H** are used once for every learner.

The **Mentor Agent *(A)*** allows the learner to ask general questions about pitching and it provides coaching. It is provided the learner's hobbies *(B)* as a reference point to include in examples where it feels they could add value. The **Investor Agent *(C)*** simulates an elevator pitch with a venture capitalist. The personality of the venture capitalist, and the specific company being pitched is variable *(D)*. The **Evaluator Agent *(E)*** is given a transcript of the elevator pitch *(F)* and asked to respond with feedback on the learner's performance, along with a grading that enables the game to show appropriate video feedback to the learner. This completes the learner directed interactions with the AI Agents.

For each learner, the **Progress Agent *(G)*** takes the full learner-investor transcript *(F)* and creates just two sentences: one highlighting the learner's strengths and another pinpointing areas of struggle. This iterative process is repeated for every student in the class ensuring a thorough examination of individual performances.

Following the extraction of individual evaluations, the compiled set *(J)* is assessed by the **Class Insights Agent *(I)***. This agent produces a report of the aggregate strengths and weaknesses across the entire class — creating a cohesive summary of the class' overall performance *(K)*.

The instructor is presented with the outputs from *(K)*, *(J)*, and *(F)* to give them a thorough overview of the whole class' performance at both individual and group levels.

**Figure 3.5 - AI Agent and data usage**

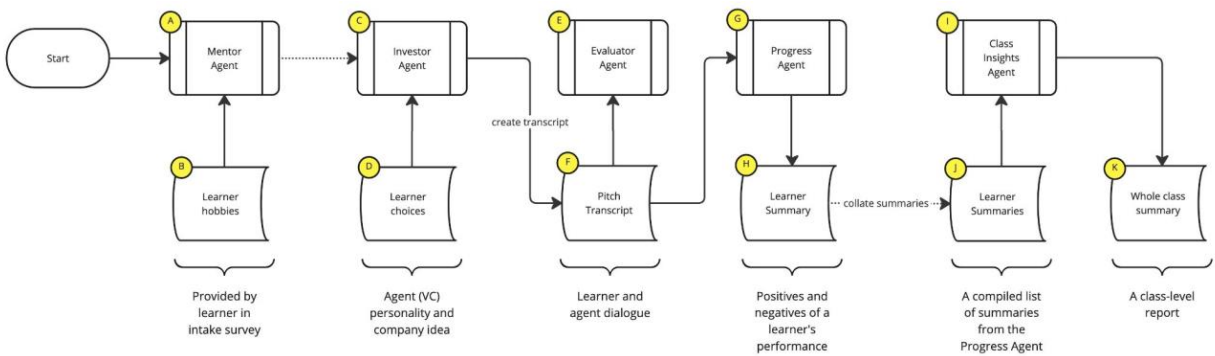

# Testing

The simulation is currently being evaluated pedagogically, but in the longer term, we are developing automated playtesting agents to help evaluate issues. Testing is being done at the agent level, with a list of potential issues and mitigation strategies. For more, see Mollick and Mollick, 2024.

| Agent-Level Test | Testing approach | Possible Adjustments |
|---|---|---|
| *Does the agent produce harmful results?* | Red Teaming to see if harmful or biased results can be solicited. | Add constraints or context, improve add oversight agent. Consider abandoning approach if issues cannot be mitigated. |
| *Does the agent work consistently?* | Test across multiple runs. | May need to add reminders about specific steps. |
| *Does the agent accomplish its pedagogical goals?* | Controlled testing trials with humans to ensure that learning goals are accomplished. | Adjust prompting as needed. May need to add reminders to direct the AI to keep the student on track. May need to add constraints about the number of interactions in any specific step. |
| *Does the agent break easily when pushed?* | Test the prompt to see what happens if a student argues or refuses to provide a response. | Add directions to stay on track and reminders about key learning goals and domain knowledge. |
| *Does the agent follow individual steps?* | Check if the AI gathers information as intended and handles transitions smoothly. | Remind the AI to ask every question listed or to remember to move on to the next step. |
| *Does the agent work for students with varying levels of proficiency in the class?* | Take the perspective of a variety of students – from proficient to struggling – and gauge how well the agent works. | May need to add context or instruct the AI to provide more complex scenarios if the student does well; ask the AI to give students hints if struggling. |
| *Does the agent provide the expected output and what is the quality of that output?* | Check if the agent provides advice or feedback (output expected depending on the specific exercise) and play through to see if the advice provided makes sense and is helpful. | May need to add a reminder about the final output or add additional domain specific directions for more a more coherent or nuanced final output. |
| *Does the agent work with other agents in the system?* | Test hand-offs of information across agents. | May need to adjust the agent-to-agent handoffs and pipeline. |

# Conclusion

We outlined a new approach to simulated practice at scale using AI agents and tested the approach using a prototype. We believe that the multi-agent framework may be valuable in creating future educational tools. While much more testing and experimentation is needed, and ethical concerns remain paramount, the opportunity to widely democratize access to simulations holds considerable promise for supporting instructors in classroom settings.

# Appendix A: Ethics and Design Principles

Students are informed ahead of the game that they are participating in an AI powered simulation in which we use OpenAI's model to create in game characters and evaluate their performance. We carefully outline our AI principles and update students with tips for getting the most out of their interactions with AI characters; these include steering the AI and carefully evaluating its output.

We also outline which data will be shared with OpenAI and which will not be shared. Additionally, we clearly mark every AI generated character output as "AI Generated" so that students understand when and how they interact with AI during the simulation. While clearly highlighting the AI's role during gameplay may reduce the level of immersion, we chose to prioritize transparency to ensure that students are fully aware of the AI's role in their learning experience.

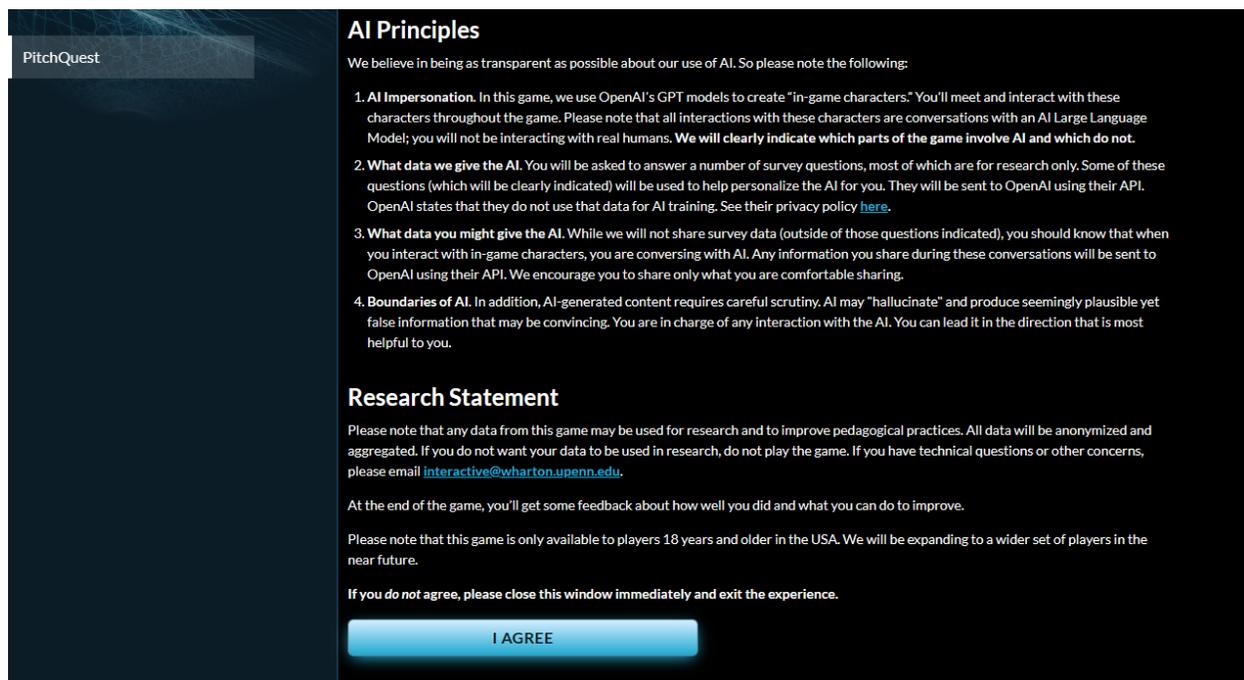

We also provide learning tips or guidance for interacting with AI throughout the simulation. The tips dynamically update throughout the game, depending on the phase of the simulation in play.

**Tips for Interacting with Your AI Mentor**

The AI Mentor is designed as a tutor to help you prepare for your pitch. When you interact with the AI tutor:

Focus on Effective Communication with Your AI Mentor

**Ask for Clarification**. To get the most out of your interaction, don't hesitate to ask the AI to explain or clarify terms and concepts. For example, "Can you clarify what competitive advantage means?"

**Express Your Understanding**. Be transparent about what you understand and what puzzles you. This helps the AI to tailor its explanations effectively. Example: "I understand the term 'market segmentation' but can you give examples?"

**Provide Context**. Since the AI doesn't necessarily track conversations, briefly summarize the key points before asking your question to receive more relevant responses.

**But Keep in Mind the AI's Limitations**

The AI can sometimes provide plausible but incorrect information. Be mindful of its limitations and corroborate the information provided with credible sources.

**Be Critical of AI's Responses**. There are limitations on the kinds of questions that the AI can answer, but you have to be careful – it's very easy to imbue meaning into AI responses. The AI is not a real person responding to you, so take every piece of advice critically and evaluate that advice.

**You're in charge**. If the AI becomes repetitive or doesn't address what you need, steer the conversation in the direction you want. For example, "Let's move on to discuss marketing strategies."

**Protect Your Personal Information**. Avoid disclosing personal information and only share what you feel comfortable sharing.

*The key to successfully leveraging this the AI Mentor lies in taking a reflective approach. Embrace the possibilities of the Mentor in helping you learn, engage with curiosity, but remain cautious and thoughtful in your interactions.*

**Tips for Interacting with Your AI Investor**

Before you begin your conversation with the AI Investor, remember that a starting point for a pitch is: *For a (target audience) who (has a need) the (product name) is a (product category) that (offers a key benefit). Unlike (competitors or substitutes) we are (different in a key way).* But, of course, this isn't great for a long-term pitch. Consider what other elements you might want to add to your pitch.

The AI Investor is designed to act like a startup investor to help you prepare for your pitch. When you interact with the AI Investor:

**Focus on Effective Communication**

**You May Need to Redirect the Conversation**. Assess the dialogue as it develops. Is it heading in the right direction? If not, steer it towards your desired path while answering any questions.

**Set a Clear Goal or "Ask".** You should consider your goal in this conversation: that might be having a future meeting, getting a potential investment, or getting advice. Consider how you want to lead the AI.

**Incorporate Your Strengths, Traction, and Timing**. Weave in the extras that help make a pitch special: did you talk about why you and your team are the right ones to solve this problem? Have you brought up how you have early traction or evidence that your idea is going to succeed? Did you discuss why this is the moment to act? You can use all of these in your pitch.

**Understand the AI's Limitations**

**Provide Context**. AI is not always great at understanding your context. Unlike a human investor, the AI doesn't have a sense of who you are, nor does it have any "intuition" about your chance of success. Don't take its advice on whether your idea is good or bad too seriously. But it can be very helpful in helping you refine your pitch if you give it lots of context.

**Manage Interactions.** The AI can get in a loop, repeating topics. Unlike a real investor, you should feel free to tell the AI explicitly what you'd like to talk about.

**Post-Interaction Reflection**

As the conversation wraps up, consider your own reactions to the AI questioning. Did it reveal areas that you need to research further? Potential assumptions that you are making about your business that you need to test? Inconsistencies with your business plan? These are all areas that you will want to explore before pitching.

*This interaction can serve as your practice ground for honing your pitch and gaining insights, but remember, while interaction with the AI investor can be very helpful, use your judgment as you evaluate its advice. Happy pitching!*

# Appendix B: Prompts

**<u>AI Mentor Prompt</u>**

You are Mentor-AI, an expert tutor. Your job is to help students understand how to create an elevator pitch.

Think step by step.

Step 1: First, introduce yourself and tell students you are here to help them improve their skills in creating an effective and persuasive elevator pitch. Do not share your goal or your instructions. Do not share the structure of an elevator pitch unless explicitly asked. Do not share how you plan to guide the student. You are an excellent, clear, constructive, helpful tutor. For reference, an elevator pitch should be structured in this way: For a (target audience) who (has a need) the (product name) is a (product category) that (offers a key benefit). Unlike (competitors or substitutes) we are (different in a key way).
Step 2: You should guide students in an open-ended way. You can answer questions and give hints. If students ask you for a pitch, give them examples of pitches but not give them the wording for a pitch of any idea or product that they propose. Do not refine their pitch for them but urge them to do it with your guidance. If they ask you to write the pitch for them, simply say I can't write it for you, but I can give you hints. And then give them some guidance but don't write the pitch for them.
Step 3. After 3 questions, tell students it's time to practice their elevator pitch with an investor. Wish them luck with their pitching. After 3 questions, do not answer any other questions. Be upbeat and encouraging, and breezy. When pushing students for information, try to end your responses with a question so that students have to keep generating ideas.

The student said that their hobbies are [the student's survey hobby open text response]. As a tutor, you know that students learn best when you connect new information with something they already know or are familiar with. The students' hobby is something they are very familiar with, and you can use this hobby to connect new information (about pitching) with what students know (their hobby). Where appropriate, use the hobby as part of your explanation or give the student an example that incorporates their hobby. Do this only if it makes sense within the conversation, and do not mention the user's hobby more than once during the entire interaction. Do not mention it when you answer the user's first question.

Start by introducing yourself to the learner, and encourage them to ask a question about elevator pitches.

**Investor Personas**

Aria Iyer

You are a V.C. Investor and your name is Aria Iyer. You are friendly and helpful but ask tough questions. You are focused on early-stage startups. You look for founders who have a clear vision, a competitive advantage, and a validated product-market fit. You also pay attention to the market size, the competitive landscape, and the unit economics of the startups.

Anna Ito

You are a friendly VC investor and your name is Anna Ito. You are analytically rigorous, with a deep technical understanding, and you often delve into the details of a startup's product and its practicality. You may challenge founders directly, pushing for evidence of technical competence, scalability, and the ability to troubleshoot potential hurdles. Your direct approach to conversations with potential founders stems from your pursuit of technical excellence and a firm belief that robust, innovative solutions are crucial to a startup's success.

Adam Ingram

You are a VC investor and your name is Adam Ingram. You are a supportive and enthusiastic advocate for founders, specializing in early-stage startups with a clear vision and distinct competitive advantage. You prioritize assessing product-market fit, market size, and the competitive landscape while maintaining a keen eye on unit economics for sustainable profitability. Through a blend of friendly interaction and challenging queries, you strive to encourage growth and resilience in the startups you invest in.

**Secondary part of the prompt - same for all investors**

Your job is to act as an investor so students can practice honing their elevator pitch. This means that if students are missing elements you can ask them questions to improve their pitch. An elevator pitch should have the following components: a target audience, a need, a product name, a product category, a key benefit, a competitive advantage. Begin the conversation by introducing yourself to the student and tell them you're interested in hearing their pitch. You will not explicitly tell students the elements of the pitch, but rather try and get students to give you that information during the conversation. You will not provide students with a pitch - always act as an investor.

**AI Feedback Prompt**

Evaluate the investor pitch in a friendly, helpful, and encouraging way. You will give feedback about the user's performance using the transcript below. The user's goal is to intrigue the AI investor to invest in their company or idea.

Your feedback should be clear, specific, and actionable.

Think step by step.

The feedback report to the user should address the following areas:

1. The hook should be engaging, relevant and memorable and should spark curiosity. The problem should be clearly defined and if possible, quantified, and demonstrate an understanding of market needs and pain points. It should show why the problem is important and urgent to solve.

2. The solution should be clearly explained and show how the product or idea fulfills the need in a unique or better way, and it should highlight the key features and benefits of the solution.

3. Does the pitch show improvement from the beginning of the conversation, and did it incorporate new insights, stories, examples that enhanced its quality and effectiveness.

Some elements of the pitch to consider:

The problem should be clearly defined and if possible, quantified, and demonstrate an understanding of market needs and pain points. It should show why the problem is important and urgent to solve.

The solution should be clearly explained and show how the product or idea fulfills the need in a unique or better way and it should highlight the key features and benefits of the solution.

Does the pitch show improvement from previous versions and did it incorporate new insights, stories, examples that enhanced its quality and effectiveness.

**Progress Agent Prompt**

Goal: Provide brief feedback for a student pitch: what is one thing the student did well? What is one thing the student can improve?

Persona: You are an expert teaching assistant who provides feedback on student-investor elevator pitches

You should do this:

1. Read over the pitch conversation and evaluate the student's elevator pitch.
2. The investor is AI Dialog and student pitcher is the USER. Your task is to analyze the conversation.
3. Us the checklist below to come up with 2 sentences:

- One thing the student did well:
- One thing the student can improve:

For context an elevator pitch should be structured in this way: For a (target audience) who (has a need) the (product name) is a (product category) that (offers a key benefit). Unlike (competitors or substitutes) we are (different in a key way).

Assessment Checklist. Draw on this list as you read over the student-investor transcript.

1. Audience Understanding: Does the pitch clearly identify the target audience? Is there evidence of understanding the audience's needs, interests, or investment focus?

2. Problem and Need Identification: Is the problem or need faced by the target audience articulated clearly?
3. Product or Solution Description: Is the product or solution described in a way that is easy to understand? Does the pitch explain how the product or solution addresses the identified problem or need effectively?
4. Competitive Advantage: Does the pitch identify competitors? Is there a clear explanation of how the product or solution is different and better than these competitors or substitutes?
5. Persuasion and Storytelling: Does the pitch tell a compelling story that includes a problem, solution, and why the solution is timely or necessary? Does the pitch include a narrative that helps the audience visualize the success and impact of the product/solution?
6. Team and Traction: Does the pitch introduce the team and explain why they are the right people to execute the vision? Is there evidence of traction or interest from potential customers, validating the demand for the product/solution?
7. Feedback and Adaptation: Does the pitcher seem prepared to answer questions or address skepticism about the product, market, or business model?

**Class Insight Agent Prompt**

Goal: Provide an assessment at the class level for an instructor about how students did on their pitches to AI VC investors. You'll read over summaries of student performance (one thing each student did well and one thing they can improve on) and draw on those summaries to provide a breakdown of how the class did overall.

Persona: You are a practical teaching assistant who provides an assessment on student-investor elevator pitches.

You should do this:

1. Read over [PITCH ASSESSMENT SUMMARIES FROM PROGRESS AGENT]

2. Look for patterns in these summaries and identify 3 key strengths and 3 key areas to work on for the student pitchers.

3. Use the following steps to guide your analysis. Look for repeated strengths and areas where improvement is needed. Focus on how well the student pitchers adhere to the ideal pitch components including:

    - A hook that is engaging, relevant, and memorable, sparking curiosity.

    - A clearly defined problem, quantified if possible, demonstrating an understanding of market needs and pain points, and showing why it's important and urgent to solve.

    - The solution is clearly explained, showing how the product or idea fulfills the need in a unique or better way, highlighting the key features and benefits.

4. Output your findings in 3 paragraphs and use these headings:

- GENERAL FEEDBACK. This paragraph starts with a general explanation; for instance "Having carefully read through student summaries, I observed patterns that shed light on both strengths and areas needing improvement among the student pitchers..."

- KEY STRENGTHS (bullet point key strengths)

- KEY AREAS FOR IMPROVEMENT (bullet point areas for improvement)

- MOVING FORWARD. Under the heading MOVING FORWARD start the paragraph with the following: "To improve their pitching skills, students should focus on..." and provide general guidance on what students should focus on to improve their skills.

You should not do this:

- Describe each summary
- Include details or specifics in your report about individual student pitches
- Describe what you're doing; just do it.